\documentclass{article}

\usepackage[dblblindworkshop, final]{neurips_2026}
\workshoptitle{NeurIPS 2026 Workshop on On-Device Intelligence: Foundation Models under Real-World Constraints.}

\usepackage[utf8]{inputenc} 
\usepackage[T1]{fontenc}    
\usepackage{hyperref}       
\usepackage{url}            
\usepackage{booktabs}       
\usepackage{amsfonts}       
\usepackage{acronym}
\usepackage{nicefrac}       
\usepackage{microtype}      
\usepackage{xcolor}         
\usepackage{graphicx}       
\usepackage{pifont}
\usepackage{booktabs,multirow,rotating,amssymb}
\newcommand{\cmark}{\ding{51}}
\newcommand{\xmark}{\ding{55}}
\usepackage{amsmath}
\title{Agentic RF Intelligence: Multi-Timescale 6G Sensing and Reasoning with On-Device Foundation Models}

\author{%
  Jaron Fontaine\\
  Department of Information Technology\\
  Ghent University, Belgium\\
  \texttt{jaron.fontaine@ugent.be} \\
  \And
  Jelle De Moerloose\\
  Department of Information Technology\\
  Ghent University, Belgium\\
  \texttt{jelle.demoerloose@ugent.be} \\
  \And
  Xander Vanparys\\
  Department of Information Technology\\
  Ghent University, Belgium\\
  \texttt{xander.vanparys@ugent.be} \\
  \And
  Anton Lambrecht\\
  Department of Information Technology\\
  Ghent University, Belgium\\
  \texttt{anton.lambrecht@ugent.be} \\
  \And
  Eli De Poorter\\
  Department of Information Technology\\
  Ghent University, Belgium\\
  \texttt{eli.depoorter@ugent.be} \\
  \And
  Adnan Shahid\\
  Department of Information Technology\\
  Ghent University, Belgium\\
  \texttt{adnan.shahid@ugent.be} \\
} 

\begin{document}

\acrodef{WPFM}{Wireless Physical-Layer Foundation Model}
\acrodef{LLM}{Large Language Model}
\acrodef{RF}{Radio Frequency}
\acrodef{MCP}{Model Context Protocol}
\acrodef{USRP}{Universal Software Radio Peripheral}

\maketitle

\begin{abstract}

Future 6G systems share a central challenge with Physical AI: combining sensing, reasoning, and action on network-edge infrastructure in rapidly changing physical environments under strict latency, compute, and energy constraints. Addressing this challenge requires multi-timescale intelligence, combining fast perception that tracks wireless phenomena within milliseconds with slower reasoning that directs sensing and adapts network policies and resources over seconds. In this paper, we present a vision for Agentic RF Intelligence based on a dual-loop architecture that decouples fast, locally autonomous wireless sensing (e.g., detecting spectrum dynamics, interference, and signal sources) from slower agentic reasoning and orchestration. We provide an on-device prototype in which the full stack runs on a single NVIDIA Jetson Thor connected to a B200 Mini USRP. The fast loop processes IQ samples using signal-processing tools and wireless physical-layer foundation models (WPFM), while a local Large Language Model (LLM) asynchronously interprets RF events, invokes tools and steers sensing. Our experiments confirm the separation in timescales, as WPFMs operate at millisecond latency, while LLM interactions take seconds to minutes. The fast loop remains operational during agentic reasoning, providing initial evidence for the feasibility of this decoupled architecture. We conclude by outlining extensions toward memory-driven self-improvement, world models, network control, and multi-agent operation.

\end{abstract}

\section{Introduction}
\begin{table}[t]
\caption{Comparison of agentic wireless systems and Wireless Physical-layer Foundation Models (WPFMs). Existing wireless agentic systems reason over predefined outputs from fixed sensing pipelines. \ac{WPFM}s offer configurable physical-layer perception but lack agentic reasoning. Our vision bridges this gap: the agent invokes on-device \ac{WPFM}s and reconfigures perception on demand.}
\label{tab}
\centering\small
\setlength{\tabcolsep}{3.5pt}
\renewcommand{\arraystretch}{0.90}
\begin{tabular}{@{}cl lllc@{}}
\toprule
& & \textbf{Perception} & \textbf{Control} & \textbf{Sensing} & \textbf{On-device} \\
\midrule
\multirow{6}{*}{\rotatebox{90}{\footnotesize Agentic}}
& \citet{ameur2026llmpoweredagenticai5g6g} & Logs, KPIs & Agent-directed & -- & \xmark \\
& \citet{ferrag20266g} & Semantic state & Agent-directed & Fixed pipeline & \xmark \\
& \citet{elkael2026agentran} & KPIs, alerts & Agent-directed & -- & \xmark \\
& \citet{li2026agentic} & CSI, AoA, Doppler, SNR & Agent-directed & Fixed pipeline & \xmark \\
& \citet{navidan2026toward} & WPFM KPIs & Agent-directed & Fixed pipeline & \xmark \\
\midrule
\multirow{3}{*}{\rotatebox{90}{\footnotesize WPFM}}
& \citet{towards_WPFM} & Raw IQ, CIR & -- & Fixed pipeline & \xmark \\
& \citet{SpectrumFM} & Raw IQ & -- & Fixed pipeline & \xmark \\
& \citet{Lightweight_WPFM} & Raw IQ, CIR & -- & Fixed pipeline & \cmark \\
\midrule
& \textbf{This work} & \textbf{Raw IQ $\rightarrow$ embeddings} & \textbf{Agent-directed} & \textbf{Agent-directed} & \cmark \\
\bottomrule
\end{tabular}
\end{table}

6G wireless systems are expected to become increasingly autonomous, moving beyond communication and reconfiguration towards intelligent systems that can sense their environment, reason about it, and decide what to do next, reducing reliance on specialized human expertise as wireless deployments become more complex and heterogeneous \citep{ferrag20266g}. In such systems, intelligence is not limited to interpreting observations. The system must decide which measurements to perform, which tools to invoke, and what to do in response. Agentic AI provides a natural framework for this behaviour through goal decomposition, tool use, memory, planning, and closed-loop adaptation, as demonstrated by recent work in wireless network control~\citep{elkael2026agentran,ameur2026llmpoweredagenticai5g6g}. 

However, bringing real-time sensing and reasoning to the wireless physical-layer is fundamentally different from applying an agent to the already-abstracted network state. Wireless events last microseconds to milliseconds: frequency hopping can occur every 625 µs and a Wi-Fi burst is over within a millisecond. Perception slower than this does not delay the observation but misses it. Language models respond in hundreds of milliseconds to seconds \citep{ferrag20266g}, and the raw signals (IQ, CSI, CIR) are too high-rate for constrained on-device buffering and too low-level for a \ac{LLM} to interpret directly, so passing them upward floods the reasoning layer rather than informing it \citep{li2026agentic}. Prior work therefore exposes higher-level information such as KPIs, estimates, or semantic state to the reasoning layer~\citep{elkael2026agentran, ferrag20266g}. This makes the information manageable for an agent, but limits what can be perceived. What is needed is not a faster reasoner, but an intelligent perception layer that keeps pace with the sensing signals while remaining open to what the agent asks of it.

A perception layer of this kind is increasingly within reach, with recent advances in \ac{WPFM} providing a promising foundation. \acp{WPFM} can operate directly on heterogeneous low-level wireless data and, after task-specific retraining or fine-tuning, support diverse downstream tasks, such as technology and modulation recognition, interference or jamming detection, and human activity recognition \citep{towards_WPFM,WPFM_mohammed,SpectrumFM}. Lightweight variants can achieve inference times under a millisecond on edge hardware \citep{Lightweight_WPFM}, showing that learned perception can keep pace with rapidly changing wireless signals, while leaving higher-level reasoning and decision-making to operate asynchronously. However, \acp{WPFM} and agentic wireless systems have largely been studied separately. \citet{navidan2026toward} combine them conceptually, but the agent only consumes the \ac{WPFM}'s performance metrics to assess the network configuration. It never invokes the \ac{WPFM} as a tool for a downstream task of its own choosing. \acp{WPFM} should instead be used as a configurable model to the wireless observation itself, complementing deterministic signal processing and reconfigurable sensing functions. This would provide reusable embeddings rather than task-specific outputs, enabling the agent to address questions that were not anticipated during model design. The open challenge is therefore how to connect this intelligent perception layer with agentic decision-making and \ac{RF} sensing tools in a single on-device system, allowing an agent to select, configure, and combine sensing capabilities as needed under the compute, memory, power, and timing constraints of the target hardware and application~\citep{liang2026foundation, 11643535}. Table~\ref{tab} summarizes this shift from agentic network control and fixed physical-layer inference to agentic on-device sensing and control.

In the remainder of this paper, we propose a dual-loop vision for agentic \ac{RF} sensing and discuss the key challenges of realizing it on-device at network-edge infrastructure. We then take a first step towards this vision with \texttt{Agentic-USRP}, a proof-of-concept running on an NVIDIA Jetson Thor with a B200 Mini \ac{USRP}, and evaluate \acp{WPFM} for perception and open-source LLMs for agentic reasoning on this platform. Finally, we discuss extensions towards wireless world models, simulation, and multi-device and multi-agent systems.

\section{Agentic RF Framework: Dual-Loop Architecture and On-Device Challenges}
\label{sct:architecture}

\begin{figure}[th]
    \centering
    \includegraphics[width=1\linewidth]{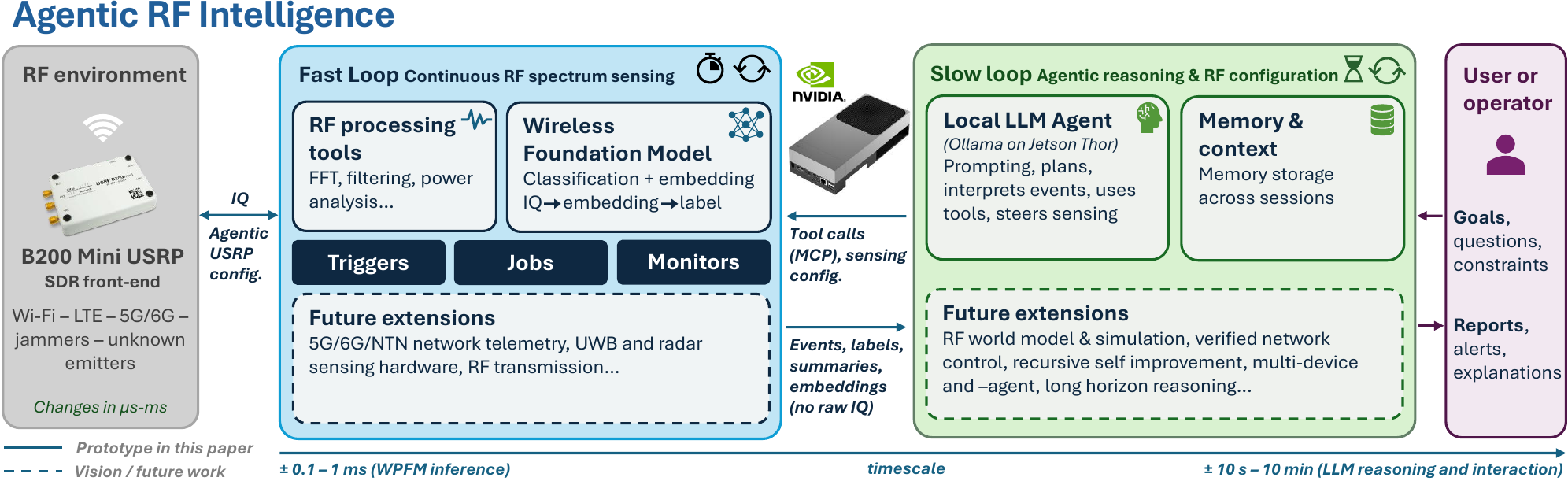}
    \caption{An Agentic AI Framework for real-time on-device wireless sensing, coupling fast RF sensing with slower agentic reasoning. Solid/dashed lines indicate the prototype and future extensions.}
    \label{fig:framework}
\end{figure}

%
Our proposed vision and framework for agentic AI in wireless networks, illustrated in Figure 1, is organized around two tightly coupled but fundamentally different execution loops deployed on network-edge infrastructure, with the agent as the reasoning and orchestration component.

\textbf{The fast loop}, shown on the left side of Figure \ref{fig:framework}, operates close to the physical interface and continuously processes heterogeneous observations such as IQ samples, CSI, CIR, Doppler, or spectrum measurements at a high rate. Signal-processing tools extract measurements, while WPFMs produce compact representations that support time-critical downstream tasks. Under strict latency, reliability, and power constraints, these capabilities can be exposed as tools to the agent through an interface such as the \ac{MCP}, on USRPs, Ultra Wideband devices, or other network-edge infrastructure, including distributed processing elements in future 6G Radio Access Networks. The fast loop monitors these measurements and triggers on relevant changes, such as signals, technologies, modulations, or anomalies appearing. Triggered events are asynchronously reported to the slow loop, giving the agent a compact view of the RF environment.


\textbf{The slow loop}, shown on the right side of Figure \ref{fig:framework}, contains an agent for long-horizon reasoning and orchestration. It combines perceptual state from the fast loop with memory, external knowledge, and potentially an RF World Model to reason about evolving wireless conditions and select appropriate actions. Through its tool interface, the agent can invoke fast-loop capabilities, such as measurements, signal analysis, WPFM inference, and sensing configuration, as well as slower expert tools such as memory retrieval and simulation. When the fast loop reports an unknown emitter, for example, the agent may request a modulation classification from the WPFM, check memory for earlier sightings, and narrow the sensing band, all without ever entering the real-time path.

The main benefit of this framework is that it closes the loop between perception and reasoning: the agent steers what the fast loop observes, while those observations guide subsequent agent decisions. Together, these components form an agentic harness connecting the agent to perception, expert tools and wireless actions for wireless intelligence. However, realizing this interaction on-device introduces challenges particularly pronounced in wireless systems, several of which \cite{li2026agentic} also identify:



\begin{enumerate}
    \item \textbf{Timescale mismatch:} Wireless channels can change within microseconds or milliseconds, while agentic reasoning, retrieval, simulation, and tool use may take much longer. The fast loop must therefore retain local autonomy, with slow-loop decisions applied asynchronously. Systems must also determine whether delayed decisions remain valid.

    \item \textbf{Resource contention:} Fast- and slow-loop processing compete for limited on-device compute, memory, and power resources. Agentic reasoning or tool execution therefore requires resource isolation and prioritization to protect latency-critical wireless processing.

    \item \textbf{Representation and context mismatch:} Raw physical-layer data is too large and volatile for direct agent processing, while excessive compression can hide rare events or distribution shifts. The interface between loops must therefore balance physical fidelity, contextual relevance, and on-device resource constraints, exposing physically grounded summaries rather than raw representations that the reasoner cannot reliably interpret.

    \item \textbf{Action safety and verification:} Agentic actions can directly modify sensing and communication behaviour through tools that configure wireless hardware or system parameters. Actions must therefore be validated against physical, operational, and real-time constraints before execution, while allowing the agent sufficient flexibility to adapt to unseen conditions.

\end{enumerate}

We therefore view the fast/slow-loop boundary and its interaction with agents as a key research problem for on-device agentic intelligence in wireless systems. Future systems must jointly determine what should be processed locally, what should be abstracted for agentic reasoning, which tools and actions should be exposed to the agent, and when slow-loop intervention is worthwhile without violating real-time operation. Addressing this interaction is essential for moving agentic AI from high-level orchestration toward physically grounded, resource-aware, and adaptive wireless intelligence.


\section{An On-Device Agentic RF Prototype}
\label{sec:implementation}

To provide an initial validation of the proposed fast/slow-loop interaction, we implemented a proof-of-concept agentic harness, \texttt{Agentic-USRP}, on an NVIDIA Jetson Thor coupled to a B200 Mini USRP (which supports real-time sensing over 56 MHz bandwidth). The PoC demonstrates how an agent can use on-device RF perception to answer higher-level sensing questions and adapt its measurements accordingly. For example, given a request such as \textit{“Our wireless link is dropping. Is RF interference responsible, and should we track it?”}, the agent can adaptively survey the spectrum, characterize the signal, compare it against current and stored observations, and initiate continued monitoring when warranted. Appendix \ref{app:execution_trace} shows a representative trace of this interaction pattern, by illustrating tool orchestration rather than a validated diagnosis. The design is deliberately compact, allowing this study to focus on the loop boundary, tool interaction, and on-device model scale.

\textbf{The fast loop} is implemented as a background worker that controls the radio and supports three operations: (i) \textbf{Triggers} periodically evaluate trigger conditions and notify the agent when those conditions are met \textit{(e.g., “Alert me if channel occupancy exceeds 50\%.”)}; (ii) \textbf{Jobs} perform a requested sensing task once and return the result \textit{(e.g., “Classify the signal at 2.437 GHz.”)}; and (iii) \textbf{Monitors} observe a frequency band for a fixed period and return one aggregated report \textit{(e.g.,  “Observe this band for 60 seconds and summarize its activity.”)}

The agent configures these operations through the tool interface, while their execution remains entirely within the fast loop. When a trigger fires, a job completes, or a monitor ends, the fast loop publishes an event to a thread-safe, sequence-ordered event bus. The slow loop retrieves these notifications asynchronously, while the fast loop never waits for the LLM, decoupling radio operation from reasoning latency. Each event carries a timestamp and policies expire automatically, however, the prototype does not yet re-check that the triggering signal is still present before the agent acts on it.

\textbf{The slow loop} uses a local LLM through Ollama (with Q4\_K\_M quantization), paired with memory that stores observations, conclusions, and hypotheses across sessions. The agent accesses RF sensing, signal analysis, memory, and fast-loop control through a common FastMCP interface. The WPFM is exposed as one of these fast-loop capabilities: the agent selects the observation target and downstream task, while the fast loop computes embeddings on-device and returns a task label or compact summary. Raw embeddings can be retained in memory rather than exposed to the LLM, addressing the representation and context mismatch discussed in Section~\ref{sct:architecture}.

\textbf{Action safety and resource sharing:}
The LLM neither accesses the radio directly nor generates executable code, but submits validated policies and bounded tool calls. Multiple sensing tasks may remain active, but radio access is serialized, with the fast loop executing them sequentially. Agent decisions are applied asynchronously, without interrupting ongoing sensing, although this does not yet guarantee resource isolation. Agent-initiated measurements can delay acquisitions when GPU, memory, and power are not reserved between LLM decoding and fast-loop inference.

\textbf{Evaluation:}
The prototype is evaluated over the air using the B200 Mini USRP. We assess its ability to process wireless observations, reason over them, and invoke sensing and analysis actions on-device. A representative WPFM architecture is benchmarked for inference latency and throughput, while local LLMs are evaluated by reasoning length, end-to-end response latency, and generation throughput on the Jetson Thor.

\section{Preliminary Results and Insights}

\begin{table}[t]
\centering
\caption{Fast-loop perception results of a CNN and WPFM on the Jetson Thor. The fast-loop covers the full USRP-to-WPFM pipeline. $\dagger$ WPFM results reproduced from \cite{WPFM_mohammed}.}
\label{tab:wpfm_results}
\footnotesize
\setlength{\tabcolsep}{3pt}
\renewcommand{\arraystretch}{0.9}
\begin{tabular}{lrrrr}
\toprule
\textbf{Model} & \textbf{Params} & \textbf{Latency} & \textbf{FP32} & \textbf{FP16} \\
      &       & \textbf{(ms)   } & \textbf{(windows/s)} & \textbf{(windows/s)} \\
\midrule
CNN & 0.82M & 0.42 & 59,349 & 89,987 \\
Transformer WPFM$^{\dagger}$ & 2.91M & 1.36 & 29,441 & 47,266 \\
\midrule
\multicolumn{4}{l}{Proposed fast-loop USRP-WPFM pipeline} & 56 Msps \\
\bottomrule
\end{tabular}
\end{table}

\begin{table}[t]
\centering
\caption{Slow-loop LLM results on the Jetson Thor across three RF sensing scenarios (one run each). Mean latency and tokens/s are reported with and without reasoning. Success means the agent used the required tools and gave an answer consistent with the measured data (no over-the-air ground truth).}
\label{tab:results}
\footnotesize
\setlength{\tabcolsep}{3pt}
\renewcommand{\arraystretch}{0.9}
\begin{tabular}{lcccccc}
\hline
& \multicolumn{3}{c}{\textbf{Reasoning}} 
& \multicolumn{3}{c}{\textbf{No reasoning}} \\
\cline{2-4} \cline{5-7}
\textbf{Model} & \textbf{Success} & \textbf{Lat. (s)} & \textbf{Tok/s}
& \textbf{Success} & \textbf{Lat. (s)} & \textbf{Tok/s} \\
\hline
Gemma 4 12B
& 1/3 & 843.1 & 20.0
& 3/3 & 24.4 & 11.2 \\

Gemma 4 31B
& 3/3 & 195.7 & 7.5
& 3/3 & 43.3 & 4.8 \\

Qwen 3.5 9B
& 3/3 & 77.0 & 22.4
& 3/3 & \textbf{18.9} & 20.3 \\

Qwen 3.8 27B
& 3/3 & 90.0 & 12.1
& 3/3 & 52.7 & 11.8 \\

Nemotron 3.5 30B-A3B
& 3/3 & \textbf{37.0} & \textbf{37.8}
& 3/3 & 24.0 & \textbf{29.6} \\

\hline
\end{tabular}
\end{table}
First, we benchmark representative WPFMs to determine whether learned sensing can operate at fast-loop timescales. Table~\ref{tab:wpfm_results} summarizes the on-device latency and sustained throughput of these WPFMs. A compact CNN (0.82M parameters) achieves 0.42~ms single-window ($=2048$ IQ samples) inference latency, while 2.91M-parameter transformer encoders require 1.36~ms to perform wireless interference classifications. With FP32, FP16 quantization and batching ($BS=256$), the transformer reaches 29.4k and 47.2k inference windows/s, respectively. The complete live USRP-to-WPFM pipeline, including capture, data transfer, buffering, and inference, sustains up to 56\,Msps running in isolation (without concurrent LLM decoding). This indicates that learned perception can operate at fast-loop timescales, while data movement, buffering, and quantization remain end-to-end bottlenecks.

Next, we evaluate whether local on-device LLMs can act as the slow reasoning loop while interacting with the real-time sensing system. Five local LLMs are evaluated using over-the-air B200 Mini USRP measurements on a Jetson Thor across three increasingly agentic scenarios: \emph{A)} spectrum observation, requiring interpretation of a band scan; \emph{B)} signal investigation, requiring selection of additional tools to characterize an unexplained signal; and \emph{C)} asynchronous event investigation, in which the fast loop captures an RF event while the LLM reasons, after which the agent consumes the queued event and performs follow-up measurements. Each scenario is evaluated with and without explicit reasoning.

Table~\ref{tab:results} shows that all five models completed the asynchronous event scenario, consuming a fast-loop event and performing follow-up measurements on-device. With reasoning enabled, \texttt{Nemotron 3.5 30B-A3B (lightning)} achieved the lowest mean latency (37.0 s) and highest throughput (37.8 tokens/s), while \texttt{qwen3.5:9b} was fastest without reasoning (18.9 s). Disabling reasoning reduced latency for every model. \texttt{gemma4:12b} completed only Scenario~C among the three reasoning-enabled scenarios; its open-ended reasoning in Scenarios~A and B produced excessively long, unterminated traces, highlighting the importance of bounded reasoning for practical on-device agents.

Together, the measurements reveal a pronounced separation between sub-millisecond to millisecond physical-layer inference and LLM tool interactions lasting seconds to minutes, a latency gap of roughly 4--6 orders of magnitude. The agent is therefore suited to operator-timescale investigation rather than real-time radio control, while the fast loop continues processing as the agent reasons and steers perception. This provides initial evidence that on-device LLMs can support higher-level wireless reasoning without blocking radio operation, while the gap between isolated and live pipeline throughput necessitates an efficient dual-loop architecture with careful resource management.

Quantifying the benefits of this design is left for future evaluation, including comparisons against scripted and rule-based baselines, repeated trials with ground-truth diagnoses, and measurements of capture-to-event latency, missed events, and memory and power use under concurrent LLM load. Checking that a queued event is still valid before acting on it is a straightforward extension. Beyond prior accuracy and generalization gains~\citep{WPFM_mohammed}, it remains to assess whether WPFMs provide on-device benefits over task-specific models, e.g., supporting multiple agent-selected tasks or continual learning from agent-labeled observations.

\section{Conclusions \& Future Work}
In this paper, we presented Agentic RF Intelligence, a vision and dual-loop framework separating fast, locally autonomous physical-layer perception from slower on-device reasoning. Our \texttt{Agentic-USRP} prototype provides initial evidence for the core premise that learned perception can operate in the fast loop and be invoked or steered by an agent without placing reasoning on the real-time path. Asynchronous execution allows the fast loop to remain operational while the agent reasons and adapts subsequent sensing, and the results highlight the importance of end-to-end pipeline design beyond model inference. Future work should close the loop from memory to recursive self-improvement, where accumulated observations and conclusions become labeled cases that fine-tune the WPFM and refine the agent's own sensing strategies, with an RF world model built on tools such as NVIDIA Sionna for counterfactual simulation before committing the radio, with network control through validated and traceable actions, and with multi-agent operation across a hierarchy of sensing nodes. Together these define research directions for autonomous, on-device wireless agentic intelligence.

\bibliographystyle{plainnat}
\bibliography{references}

\newpage
\appendix

\section{Appendix: Representative Agentic-USRP Execution Trace}
\label{app:execution_trace}

To illustrate the interaction between the agent and the RF sensing
capabilities of \texttt{Agentic-USRP}, we provide a representative execution
trace for a user asking the system to investigate interference affecting a
wireless link. The trace shows the agent iteratively invoking RF sensing tools,
using their observations to guide subsequent measurements, and eventually
delegating background monitoring to the fast loop. 
The trace illustrates the interaction pattern, not a verified diagnosis. We discuss its evidential gaps at the end of this section.

For readability, the trace has been manually trimmed: repetitive UHD
initialization and sample-rate configuration messages, as well as other
low-level system logs, are omitted. We also include only short summaries of
the agent's reasoning rather than its complete internal reasoning trace.

\begin{verbatim}
[INFO] [B200] Detected Device: B200mini
[INFO] [B200] Operating over USB 3.
[INFO] RF backend: USRP hardware (B200mini)
[INFO] inference: cuda, LLM: qwen3.8
[INFO] fast loop started (interval=2.0s)

[[Slow loop]]
User prompt:
  "Our wireless link is dropping. Is RF interference responsible,
   and should we track it?"

Agent:
  I will first survey the spectrum to establish situational
  awareness, then investigate suspicious activity with targeted
  RF tools.

[[Fast loop]]
-> survey_known_bands()

[OK] 7 active band(s)
[[Slow loop]]
Result:
  2.4 GHz ISM: active, peak SNR 14.8 dB, occupancy 14%
  5 GHz WiFi:  active, peak SNR 11.9 dB, occupancy 0.1%
  915 MHz ISM: active, peak SNR 23.95 dB, occupancy 0.4%
  868 MHz ISM: active, peak SNR 9.77 dB, occupancy 0.1%
  FM:          active, peak SNR 9.17 dB
  ADS-B:       active, peak SNR 11.95 dB
  GPS L1:      active, peak SNR 10.66 dB

Agent reasoning (summary):
  The 2.4 GHz band is highly occupied, making it a plausible
  source of interference. I will check for jamming and anomalies.

[[Fast loop]]
-> run_detector(task="jamming_detect", gain_db=20)

[OK] task 'jamming_detect': 0 hit(s)
-> run_detector(task="anomaly_detect", gain_db=20)

[OK] task 'anomaly_detect': 1 hit(s)
[[Slow loop]]
Result:
  center frequency: 2487.5 MHz
  SNR:              ~14.2 dB
  reported bandwidth: ~18.3 kHz

Agent reasoning (summary):
  The anomaly lies just above the 2.4 GHz ISM band edge.
  I will zoom in to characterize the emission.

[[Fast loop]]
-> zoom_in(center_freq_mhz=2487.5, span_mhz=4,
           bandwidth_method="occupied")

[OK] wideband_signal
  occupied bandwidth: ~3.7 MHz
  SNR: ~11.6 dB
  channel-plan match: none
-> classify_modulation(center_freq_mhz=2487.5, bandwidth_mhz=4)

[OK] modulation classification
  envelope: amplitude_varying
  top hypothesis: 64QAM
[[Slow loop]]
Agent reasoning (summary):
  The emission is wideband and does not match a known channel plan.
  I will continuously monitor the 2.4 GHz band to determine whether
  this activity persists and whether it correlates with the link issue.

[[Fast loop continues monitoring in the background]]
-> monitor_band(band_start_mhz=2400, band_stop_mhz=2483.5,
                duration_s=20)

[OK] monitoring armed (interval=2.0s)
-> [20 s later] monitor report ready
   10 samples, 77 signals

[[Slow loop]]
Result:
  mean occupancy: 6.2%
  peak SNR:       40.4 dB at 2426 MHz (BLE channel)
  2442--2448 MHz: multiple ~2 MHz wideband signals (WiFi)
  2480 MHz:       persistent BLE activity
  unmatched tones: none

Agent:
  The observed activity is consistent with legitimate BLE and
  WiFi traffic. No narrowband unmatched tone or jammer signature
  was detected. The link drops are therefore more consistent
  with congestion or contention than intentional interference.
\end{verbatim}

This trace highlights the intended dual-loop interaction: the slow-loop agent iteratively selects sensing operations based on previous observations, while the fast loop independently executes persistent RF monitoring. In the current prototype, the 2 s interval is the actual acquisition cadence: each acquisition records a short snapshot of each sub-band (up to 56 MHz wide), so the 20 s monitor yields ten snapshots, and signals appearing between acquisitions may be missed. This is a limitation of the current implementation rather than the dual-loop design: the continuous USRP-to-WPFM pipeline from Section~4 can process every IQ window in the tuned band and pass detected events to the agent, while the agent can determine which bands to cover and revisit. The anomaly in this trace lies just outside the monitored band, making the final explanation a hypothesis rather than a verified diagnosis, as no link-loss data was correlated with the observations. Such gaps arise from the agent's sensing decisions and could be mitigated by stronger models or by checking that conclusions only reference frequencies that were actually measured.

\end{document}